\def\Granadainst{Instituto de F\'\i sica Te\'orica y Computacional Carlos I,
Facultad de Ciencias, Universidad de Granada, Campus de Fuentenueva,
Granada 18002, Spain}

\def\IAA{Instituto de Astrof\'{\i}sica de Andaluc\'{\i}a, Apartado Postal 3004,
18080 Granada, Spain}

\def\Murcia{Departamento de Matem\'atica Aplicada, Facultad de Inform\'atica, Campus
de Espinardo, 30100 Murcia, Spain}

\def\be{\begin{equation}}
\def\ee{\end{equation}}
\def\bea{\begin{eqnarray}}
\def\eea{\end{eqnarray}}
\def\ba{\begin{array}}
\def\ea{\end{array}}

\documentclass[12pt]{article}
\usepackage{amsfonts,amssymb}

\begin{document}

%\begin{titlepage}

%\begin{flushright}
%HEP-TH/9707237
%\end{flushright}

\begin{center}
{\large {\bf Erratum to:}}\\
{\large {\bf Canonical coherent states for the relativistic harmonic oscillator}}, 
J. Math. Phys. {\bf 36}(7), 3191 (1995)
\end{center}

\bigskip
\bigskip

\centerline{V. Aldaya$^{1,2}$ and J. Guerrero$^{2,3}$}

\bigskip
\centerline{July 2003}
\bigskip

\footnotetext[1]{\IAA}
\footnotetext[2]{\Granadainst}
\footnotetext[3]{\Murcia}

\bigskip

%\begin{center}
%{\bf Abstract}
%\end{center}

%\small
%\setlength{\baselineskip}{12pt}

%\begin{list}{}{\setlength{\leftmargin}{3pc}\setlength{\rightmargin}{3pc}}
%\item 
%\end{list}

%\normalsize

\vskip 0.25cm

%PACS numbers: 02.20.Qs,\  02.40.-k,\ 03.65.-w,\ 73.40.Hm

%\setlength{\baselineskip}{14pt}

%\end{titlepage}

%\vfil\eject

%\section*{I. Introduction}

In the mentioned paper \cite{A-G} we introduced higher-order (non-polynomial), 
relativistic creation and annihilation operators, $\hat{a},\hat{a}^\dag$, with canonical 
commutation relation $\left[\hat{a},\hat{a}^\dag\right]=1$ rather than the covariant one 
$\left[\hat{z},\hat{z}^\dag\right]\approx$ energy and naturally associated with the $SL(2,R)$ group. 
The canonical (relativistic) coherent states were then defined as eigenstates of $\hat{a}$.
Also, a canonical, minimal representation was constructed in configuration space by means 
of eigenstates of a canonical position operator.

Unfortunately the expression of the operator $\hat{\kappa}$ (closely related to the energy 
operator) just after formula (18), then after formula (34), was miswritten. In fact, we printed the classical 
function of $\kappa$ in terms of the functions $z$ and $z^*$ (see eq. (2)), whereas
the correct, quantum expression is:

\[\hat{\kappa}=\frac{1}{2N}+\sqrt{(1-\frac{1}{2N})^2+\frac{2}{N}\hat{z}^\dag\hat{z}}\]

This misprint had not been detected because we always used the power series expansion
(formula (17)), which features a full independence on the (energy eigenstate) basis 
$\{|n>\}$. However, very recently, H.A. Kastrup  dealing with an analogous 
construction \cite{Kastrup} has detected the above-mentioned misprint \cite{privado}. We are very
grateful to him for pointing it out.

%\section*{Acknowledgments}

\end{document}